\documentclass[aps,prb,tightenlines,preprint,showpacs,floatfix]{revtex4}
\usepackage{amsfonts,amsmath,amssymb}
\usepackage[latin1]{inputenc}
\usepackage[T1]{fontenc}
\usepackage{textcomp}
\usepackage{times}
\usepackage{bm}
\usepackage[dvips]{color,graphicx}

\usepackage[dvipsnames]{pstricks}
\definecolor{nicered}{rgb}{0.75,0.0,0.0}
\begin{document}
\title{Comment on ``Quantum critical paraelectrics and the Casimir
effect in time''}
\author{H. Chamati}
\email{chamati@issp.bas.bg}
\affiliation{Institute of Solid State Physics,
Bulgarian Academy of Sciences,
72 Tzarigradsko Chauss\'ee, 1784 Sofia, Bulgaria}
\author{N. S. Tonchev}
\email{tonchev@issp.bas.bg}
\affiliation{Institute of Solid State Physics,
Bulgarian Academy of Sciences,
72 Tzarigradsko Chauss\'ee, 1784 Sofia, Bulgaria}

\begin{abstract}
At variance with the authors' statement
[L. P\'{a}lov\'{a}, P. Chandra and P. Coleman, Phys. Rev. B {\bf 79}, 075101 (2009)],
we show that the behavior of the universal scaling amplitude
of the gap function in
the phonon dispersion relation as a function of the dimensionality
$d$, obtained within a self--consistent one--loop approach,
is consistent with some previous analytical results obtained in
the framework of the $\varepsilon$--expansion in conjunction with
the field theoretic renormalization group method
[S. Sachdev, Phys. Rev. B {\bf 55}, 142 (1997)] and the exact
calculations
corresponding to the
spherical limit i.e. infinite number $N$ of the components of the order
parameter
[H. Chamati. and N. S. Tonchev, J. Phys. A: Math. Gen. {\bf 33}, 873 (2000)].
Furthermore we determine numerically the behavior
of the ``temporal'' Casimir amplitude as a function of the dimensionality $d$
between the lower and upper critical dimension and found a maximum at
$d=2.9144$. This is confirmed via an expansion near the upper
dimension $d=3$.
\end{abstract}

\maketitle

In a recent paper P\'{a}lov\'{a}, Chandra and Coleman (PCC)
\cite{palova2009} studied the quantum
paraelectric-ferroelectric phase transition in the framework of the
self-consistent one--loop approximation applied to the familiar
quantum $\varphi^4$ model that plays an important role
in the investigations of the properties of many quantum systems near
their quantum critical points.\cite{sachdev1999} After a suitable normalization of
the parameters of the model this approach is formally equivalent to
considering an $N$-component model in the spherical limit i.e.
when the number $N$ of the components of the order parameter is sent
to infinity.\cite{moshe2003}

In their equation (33) for the gap function PCC explore the
role of the temperature as a boundary effect in the ``imaginary time''.
Interpreting the inverse temperature as a finite size in this
direction, it is possible to map the theory of quantum critical
phenomena at low temperatures on the finite size scaling
theory\cite{privman1990,brankov2000}
which shaped our current understanding of the modern theory of critical
phenomena. In this context it is shown in Ref. \onlinecite{palova2009} that at a
temperature $T$ above the quantum critical point, the gap function
$\Delta(T)$ in the phonon dispersion relation scales as
$$
\frac{\Delta(T)}{T}=\alpha_{d}.
$$
The corresponding equation for the scaling amplitude $\alpha_{d}$
was solved numerically for arbitrary dimensions in the range
$1<d<3$. The bounds on $d$ are imposed by the fact that for
$d\leq 1$ no phase transition can survive,
while for $d\geq3$ one obtains a mean field critical behavior.
Close to the upper critical dimension, the presented on
FIG. 5 dependence of $\alpha_{d}(T\rightarrow 0)$ on dimensionality
$d$ exhibits a discrepancy (it is finite) with previous analytical
considerations (where it goes to zero) as $\varepsilon\rightarrow 0$
\cite{sachdev1997} and/or $N\rightarrow \infty$ \cite{chamati2000a},
as well as numerical ones for the quantum spherical model.
\cite{chamati1998a}
Comparing $\varepsilon$-results \cite{sachdev1997} and their
numerical prediction PCC suggest that this discrepancy may be attributed
to the order in which the limits $\varepsilon\rightarrow 0$ and
$N\rightarrow \infty$ are evaluated.  In this comment we
demonstrate that at variance with PCC's claims there is no
contradiction between previous analytical
\cite{sachdev1997,chamati2000a} and numerical \cite{chamati1998a}
considerations, and the analysis based on self-consistent one-loop approximation
presented in Ref. \onlinecite{palova2009}.

Our aim is to show that the numerical treatment of Eq. (42) of Ref.
\onlinecite{palova2009} is not adequate closely beneath the upper
critical dimension $d=3$.
For the sake of completeness, we will outline the main steps
of our computations. According to Eq. (33) of PCC one has:
\begin{equation}\label{ppc}
\Delta^2=\Omega_0^2+3\gamma_c \Gamma_d\int_0^\Lambda
\frac{dqq^{d-1}}{(2\pi)^d} \frac{n_B(\omega_q)}{\omega_q}
+\frac{3}{2}\gamma_c
\Gamma_d\int_0^\Lambda\frac{dqq^{d-1}}{(2\pi)^d}
\left(\frac1{\omega_q}-\frac1q\right),
\end{equation}
where $n_B(\omega)=(e^{\omega/{k_BT}}-1)^{-1}$,
$\Gamma_d^{-1}=\frac12\pi^{-d/2}\Gamma(d/2)$,
$r$ and $\gamma_c$ are model constants, and
$\omega_q=\sqrt{q^2+\Delta^2}$. The parameter $\Omega_0^2=r-r_c$
measures the distance of the quantum parameter driving the
transition from its critical value $r_c$. In the remainder we use
$k_B=1$.
Notice that we clearly separate the thermal ($T>0$) and quantum ($T=0$)
fluctuations and introduce the cutoff $\Lambda$.

Further, via the substitution $\Delta=\alpha T$ and $q=u T$ we may write
\begin{eqnarray}
\frac{(2\pi)^d}{3\gamma_c\Gamma_d}T^{3-d}
\left(\alpha^2-\frac{\Omega_0^2}{T^2}\right)&=&
\int_0^\frac{\Lambda}{T} duu^{d-1} \frac1{\sqrt{\alpha^2+u^2}
\left[\exp\left(\sqrt{\alpha^2+u^2}\right)-1\right]}\nonumber\\
& &+\frac12 \int_0^\frac{\Lambda}{T}du u^{d-1}
\left(\frac1{\sqrt{\alpha^2+u^2}}-\frac1u\right).
\end{eqnarray}
In the low temperature region $\frac{\Lambda}{T}\gg1$,
the cutoff in the first integral can be entirely removed neglecting
exponentially small corrections. The last integral is convergent in the
ultraviolet in dimensions $1<d<3$. It may be computed extending the
integration over $u$ up to infinity to get
\begin{equation}\label{self}
(4\pi)^{(d+1)/2}
\left(\frac{\alpha^2T^{3-d}}{3\gamma_c}-
\varkappa\right)=
\Gamma\left(\frac{1-d}2\right)
\alpha^{d-1}+2^d\Gamma\left(\frac{d-1}2\right)
h_{d-1}(\alpha^2),
\end{equation}
where we have introduced the scaling variable
$$
\varkappa=\frac{\Omega_0^2}{3\gamma_c}T^{-1/\nu z}
$$
with $\nu=(d-1)^{-1}$ the critical exponent measuring the
divergence of the correlation length, $\xi$, while approaching the
quantum critical point and $z=1$ the dynamical critical exponent.
The function $h_\mu(z)$ is defined via
\begin{equation}
h_\mu(z)=\frac1{\Gamma(\mu)}\int_0^\infty\frac{u^{\mu}du}{\sqrt{z+u^2}
\left[\exp\left(\sqrt{z+u^2}\right)-1\right]}.
\end{equation}
In particular one has
\begin{equation}
h_\mu(0)=\zeta(z),
\end{equation}
where $\zeta(x)$ is the Riemann zeta function.

In the vicinity of the quantum critical point, defined by
$\Omega_0=0$ and $T=0$, the term containing $\alpha^2$ in the left
hand side may be neglected as it gives only corrections to the
leading order of $\alpha$. Then the solution to Eq. (\ref{self})
has the finite temperature scaling form
\begin{equation}\label{al}
\alpha_{d}=f_d\left(\varkappa\right),
\end{equation}
with $f_d(\varkappa)$ an universal scaling function.

For the scaling form (\ref{al}) to be valid one has to require
\begin{equation}\label{K}
T^{3-d}\ll\frac{{\it A(d)}}{(\alpha_{d})^{3-d}},\quad
A(d):=\frac{3\gamma_c}{(4\pi)^{d/2}}\left|\Gamma\left(\frac{1-d}{2}\right)\right|.
\end{equation}
This inequality is an estimation of the low temperature
region and the magnitude of $\gamma_c$ where the scaling form (\ref{al}) takes place.

\begin{figure}[ht!]
\resizebox{.6\columnwidth}{!}{\includegraphics{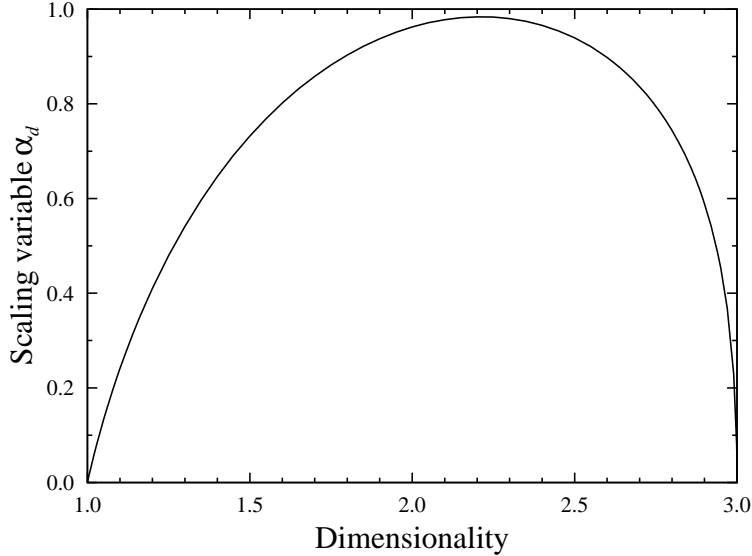}}
\caption{
Dependence of $\alpha_{d}=f_{d}(0)$ on dimensionality $d$ at
the quantum critical point i.e. $\Omega_0=0$ and $T\rightarrow 0^+$.
}
\label{alpha}
\end{figure}

The behavior of the scaling variable $\alpha_d$ above the
quantum critical point, obtained
numerically by equating the right hand side of (\ref{self}) to zero, is
presented in FIG. \ref{alpha}.
To validate our numerical results we choose to perform analytic
calculations
of $\alpha_{d}$ by considering some particular cases:
namely $d=2$, and in a close vicinity of $d=3$ and $d=1$. This leads us
to the results (see e.g. Ref. \onlinecite{chamati2000a})
\begin{equation}\label{ar}
\alpha_{d}=\left\{
\begin{array}{ll}
\pi\left(d-1\right),                     &d-1\ll1,  \\[.5cm]
2\ln\frac{1+\sqrt{5}}{2}, \qquad\qquad   &d=2,      \\[.5cm]
\sqrt{\frac{2\pi^2}3}\sqrt{3-d},         &3-d\ll1.
\end{array}
\right.
\end{equation}
One sees that the behavior of $\alpha_{d}$ for the cases
$d\rightarrow 1$ and $d=2$ seems to be correctly presented in FIG. 5
of Ref. \onlinecite{palova2009}. However both numerical (FIG.
\ref{alpha}) and
analytical computations (\ref{ar}) show that $\alpha_{d}$ vanishes
as $d\rightarrow 3$. This result disagrees with the conclusions
drawn by PCC based on the behavior presented on their FIG.
5.

In the remainder of this comment we will briefly touch on some
aspects of the so called temporal Casimir
effect or Casimir effect in time\cite{palova2009}
considered also some years ago in Ref. \onlinecite{chamati2000c} in the
framework of the quantum spherical model.

The self-consistent one-loop approximation is exact in the spherical
limit, i.e. for the theory with $N$ component order parameter
in the limit $N\rightarrow \infty$. In this case the free energy can
be computed using a variational approach based on the
Hubbard--Stratonovich decoupling
technique\cite{moshe2003} of the $P^4$ term ($P$ being
$N-$component field) in the model of
Ref. \onlinecite{palova2009}. Following Ref.
\onlinecite{chamati2000a}
we end up with an expression for the free energy per particle and per
component
\begin{equation}\label{fe}
\mathcal{F}_d(T)=\min_{\Delta}\left\{-\frac1{4\gamma_c}(\Delta^2-r)^2
+T\int\frac{d^d\bm{q}}{(2\pi)^d}
\ln\left[2\sinh\left(\frac1{2T}\omega_{\bm{q}}\right)\right]\right\}.
\end{equation}
The equation minimizing the free energy (\ref{fe}) is identical to
the self consistency
equation (\ref{ppc}) up to
the substitution $\gamma_{c}\rightarrow 3\gamma_{c}$. Notice that
expression (\ref{fe}) for the free energy can be split into a
``pure quantum'' free energy and a ``finite temperature'' contribution as:
\begin{equation}
\mathcal{F}_d(T)=\min_{\Delta}\left\{
-\frac1{4\gamma_c}\left(\Delta^2-r\right)^2
+\frac12\int\frac{d^d\bm{q}}{(2\pi)^d}\omega_{\bm{q}}
+T\int\frac{d^d\bm{q}}{(2\pi)^d}\ln\left(1-e^{-\omega_{\bm{q}}/T}\right)\right\}.
\end{equation}

Applying the assumption of Ref. \onlinecite{privman1984} for the
singular part of the free energy for classical systems within
the theory of finite size scaling to the case
of quantum critical phenomena at finite temperature, the
singular part of the free energy should
scale like
\begin{equation}\label{fss}
\mathcal{F}_d^\mathrm{sing.}(T)\sim T^{1+d/z}\mathfrak{F}\left[(r-r_c)T^{-\nu z}\right],
\end{equation}
where the function
$\mathfrak{F}$ is an universal scaling function, whose value at the
quantum critical point is equivalent to the Casimir
amplitude\cite{brankov2000} for temperature driven phase transitions
in films. Such an idea, considering the inverse temperature as an
additional dimension, has
been developed for quantum critical points in Ref.
\onlinecite{chamati2000c}. Notice however that for quantum
systems this quantity may be measured experimentally since it
is related to the amplitude of the specific heat of the system at
finite temperature.

For the model under consideration in the interval of interest
i.e. dimensions $1<d<3$ and in the vicinity of the quantum critical
point, the singular part of the free energy takes the scaling form
\begin{equation}\label{scf}
\mathcal{F}_d^\mathrm{sing.}(T)=T^{1+d}g_d(\varkappa),
\end{equation}
where
\begin{equation}\label{scaling}
g_d(\varkappa)=\frac12\varkappa\alpha_d^2-\frac12(4\pi)^{-(d+1)/2}
\Gamma\left(-\frac{d+1}2\right)\alpha_d^{d+1}
-\pi^{-(d+1)/2}\Gamma\left(\frac{d+1}2\right)h_{d+1}(\alpha_d^2)
\end{equation}
is an universal scaling function with $\alpha_d$ the solution
(\ref{al}) to Eq. (\ref{self}). It is worth mentioning that
the scaling form (\ref{scf}) is in agreement
with the scaling ansatz (\ref{fss}).

\begin{figure}[ht!]
\resizebox{0.6\columnwidth}{!}{\includegraphics{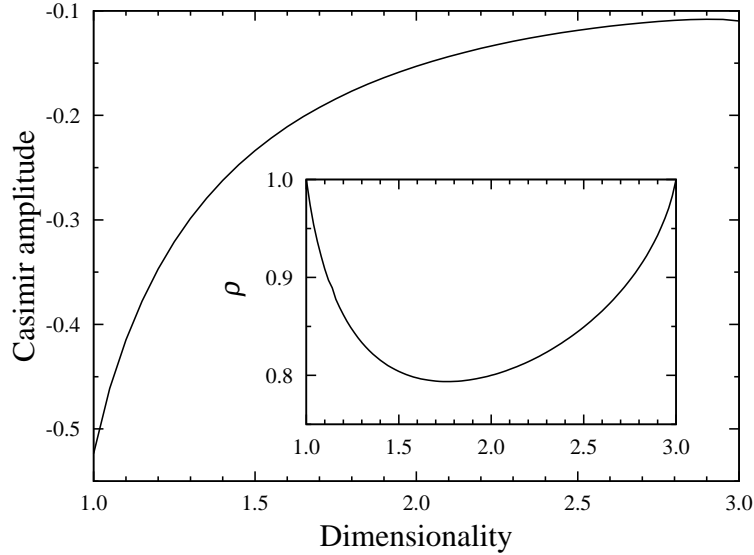}}
\caption{
The behavior of the Casimir amplitude $g_d(0)$ from Eq.
(\ref{scaling}). In the inset we graph the ratio $\rho$ defined in Eq.
(\ref{rho}).}
\label{casimir}
\end{figure}

The behavior of the scaling function (\ref{scaling}) for $d=2$ is well known
in the literature.\cite{chubukov1994,chamati2000c} Here we wish to check
its dependence upon the dimensionality. We are primarily interested in the behavior of the
amplitude $g_d(0)$ of the free energy at the quantum critical point,
as this is tightly related to the
"Casimir effect in time".\cite{chamati2000c}
For the particular case $d=2$, it can be
computed analytically resulting in\cite{chamati2000c}
\begin{equation}
g_2(0)=-\frac{2\zeta(3)}{5\pi}.
\end{equation}
For arbitrary $d$ the
behavior of $g_d(0)$ can be obtained by numerical means. This
is graphed in FIG. \ref{casimir}.
It is found that the scaling function has a maximum at $d=2.9144$.
We check our numerical results in the vicinity of $d=3$,
using an $\varepsilon=3-d$ expansion, taking into account the small
$\varepsilon$ behavior of $\alpha_{3-\varepsilon}$,
to get
\begin{equation}\label{caseps}
g_{3-\varepsilon}(0)=-\frac{\pi^2}{90}+
\left[\frac{\pi^2}{360}(7-2\gamma-2\ln\pi)+\frac{\zeta'(4)}{\pi^2}\right]\varepsilon
-\frac{\pi^2}{9\sqrt{6}}\varepsilon^{3/2}+o(\varepsilon^{3/2}),
\end{equation}
where $\gamma=0.5772$ is the Euler--Mascheroni constant. Indeed, expression
(\ref{caseps}) exhibits a maximum at $d=2.9818$.
This maximum is shifted compared to the one obtained by numerical
means due to the used approximation for $\varepsilon$.

At the borderlines $d\to1^-$
and $d\to3^+$, the values of $g_d(0)$ coincide with their counterparts of the Gaussian
theory corresponding to $\alpha_d=0$ i.e.
$$
g_d^\mathrm{Gaus.}(0)=-\pi^{-(d+1)/2}\Gamma\left(\frac{d+1}2\right)
\zeta(d+1)
$$
This can be seen in the inset of FIG. \ref{casimir}, where we present
the behavior of
\begin{equation}\label{rho}
\rho=g_d(0)/g_d^\mathrm{Gauss.}(0).
\end{equation}
which is related to the Zamolodchikov $C$-function extended to arbitrary
dimensions and nonzero temperature (see Ref. \onlinecite{brankov2000} and
references therein).
Here, we would like to point out the
similarity between the behavior of $\rho$ in FIG. \ref{casimir} and
the one obtained in Ref. \onlinecite{petkou1998} for the characteristic parameter of
the corresponding conformal field theory in dimensions $2<d<4$.

Let us note, before closing this comment, that the gap equation [Eq. (\ref{self})] with
l.h.s. equals zero is equivalent to the spherical constraint
imposed on the quantum spherical model, see
Refs. \onlinecite{chamati1998a,chamati2000c,oliveira2006} where
the finite-temperature scaling was studied in great details.

The consideration outlined in this comment remains valid
also for systems with
film geometry under periodic boundary conditions
with temperature driven phase transition. This is due to the fact that the thickness
of films is in some sense equivalent to the inverse
temperature in a quantum system.
This is a facet of the property known
as \textit{temperature inversion symmetry}
discussed in the literature that leads
to explicit conversion from Casimir force to Planck's law of
radiation. \cite{ravndal1989,fukushima2001}
Very recently, in Refs. \onlinecite{grueneberg2008} and
\onlinecite{diehl2009}, using the $\varepsilon$
expansion and/or the limit $N\to\infty$ in the framework of the
classical $O(N)$ symmetric $\varphi^4$ model with film geometry one
obtains results for the amplitude of the correlation length and
quantities related to the Casimir effect in a close
vicinity of the upper critical dimension that can be conversed to the
field of quantum paraelectric-ferroelectric phase transitions in particular and to
quantum critical phenomena in general.

This work was supported by the Bulgarian Fund for Scientific
Research Grant No. F-1517 (H.C.) and Grant No. BYX-308/2007 (N.T.).

\end{document}